\documentclass[letterpaper]{article} 
\usepackage{aaai2026}  
\usepackage{times}  
\usepackage{helvet}  
\usepackage{courier}  
\usepackage[hyphens]{url}  
\usepackage{graphicx} 
\usepackage{natbib}  
\usepackage{caption} 
\usepackage{algorithm}
\usepackage{algorithmic}
\usepackage{bibentry}

\usepackage{newfloat}
\usepackage{listings}
\DeclareCaptionStyle{ruled}{labelfont=normalfont,labelsep=colon,strut=off} 
\floatstyle{ruled}
\newfloat{listing}{tb}{lst}{}
\floatname{listing}{Listing}

\title{Agentic AI for Software: thoughts from Software Engineering community}
\author{
    Abhik Roychoudhury\textsuperscript{\rm 1}\thanks{Full-time involvement as Professor at NUS, while being Senior Advisor at SonarSource SA.
    }
}
\affiliations{
    \textsuperscript{\rm 1}National University of Singapore\\
    {\tt abhik@nus.edu.sg}
}

\begin{document}

\maketitle

\begin{abstract}

AI agents have recently shown significant promise in software engineering. Much public attention has been transfixed on the topic of code generation from Large Language Models (LLMs) via a prompt. However, software engineering is much more than programming, and AI agents go far beyond instructions given by a prompt. 

At the {\em code level}, common software tasks include code generation, testing, and program repair. {\em Design level} software tasks may include architecture exploration, requirements understanding, and requirements enforcement at the code level. Each of these software tasks involves micro-decisions which can be taken autonomously by an AI agent, aided by program analysis tools. This creates the vision of an AI software engineer, where the AI agent can be seen as a member of a development team. 

Conceptually, the key to successfully developing trustworthy agentic AI-based software workflows will be to resolve the core difficulty in software engineering - the deciphering and clarification of developer intent. Specification inference, or deciphering the intent, thus lies at the heart of many software tasks, including software maintenance and program repair. A successful deployment of agentic technology into software engineering would involve making conceptual progress in such intent inference via agents. We discuss, to some length, the AutoCodeRover agent which embodies such intent inference. The agent has been integrated into the SonarQube static analysis tool.

Trusting the AI agent becomes a key aspect, as software engineering becomes more automated. Higher automation also leads to higher volume of code being automatically generated, and then integrated into code-bases. Thus to deal with this explosion, an emerging direction is AI-based verification and validation (V \& V) of AI generated code. We posit that agentic software workflows in future will include such AI-based V\&V. 
\end{abstract}

\begin{links}
    \link{Code}{N.A.}
    \link{Datasets}{SWE-bench}
    \link{Extended version}{N.A.}
\end{links}

\section{Historical Retrospective}
Since the 1970's when programming in modern existing programming languages firmly took root - the focus has been on {\em programming in the large}. The start of this period can be traced back to around fifty years ago - when the first {\em Hello World} program was written by Brian Kerninghan in 1972-3 in the B programming language, which is the predecessor of C. Since the writing of the first "Hello World" program and the initial seminal innovations, there have been only two really major shifts in the software industry affecting {\em both the technology as well as the business of software}. 

The first shift came 25 years after the "Hello World" program, in the late nineties. Subsequent to the birth of the internet - the software delivery model as a service took root. This led to many Software-as-a-Service (SaaS) offerings in vertical domains such as customer relationship management (SalesForce being an example company in the domain). The movement from in-house software development to cloud-based hosting of software also led to movement of storage or documents to the cloud - allowing for conveniences such as collaboratively working on documents. 

In 2025, we are now at the second major shift which affects both the software technology and business- the arrival of {\em agentic AI}. With the use of agentic AI in software - the change is beyond the hosting of the software - instead, the engineering of the software itself is automated to a large degree. Since there is greater automation - this eases the writing of code fragments capturing different functionality. However, the correctness of the code and the overall trust in the agent output becomes a key concern that needs addressing. Thus, perhaps for the first time since the writing of the "Hello World" program - we face the central issue of {\em programming with trust}, instead of programming in the large.

\begin{figure*}
    \centering
   \includegraphics[width=0.9\linewidth]{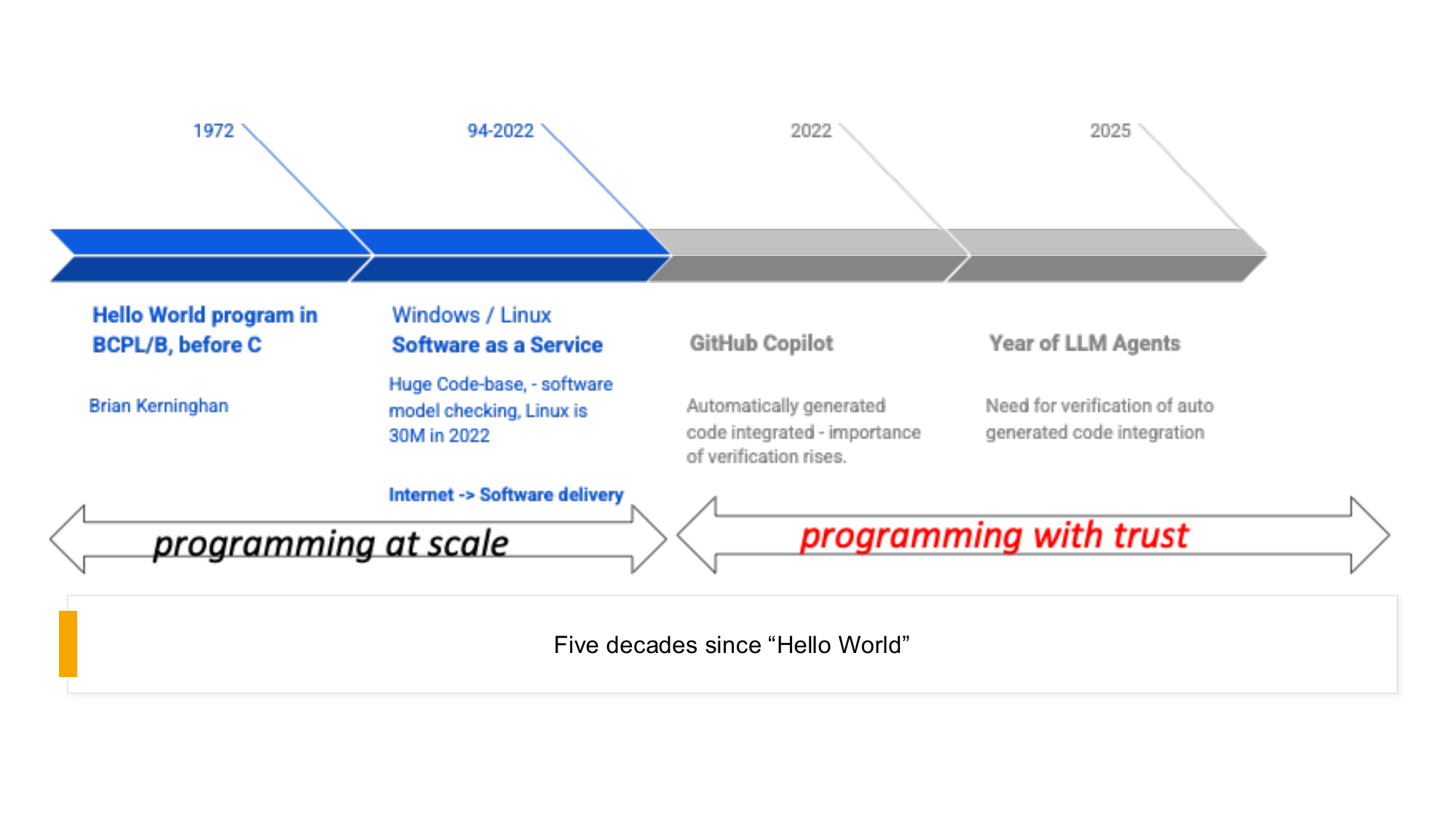}
   \vspace*{-0.4in}
    \caption{\em Arrival of Agentic AI in Software - more than fifty years after "Hello World".}
    \label{fig:history}
\end{figure*}

Figure \ref{fig:history} shows the timeline of some key developments in the software industry in the past fifty years- highlighting the two major developments: SaaS and agentic AI.  It also shows the contrast between "programming in the large" and "programming with trust". 

In the rest of the article we will navigate this tension between scale and trust, and also transcend the issues in programming to the core concerns in software engineering. We will also discuss how this tension betwen scale and trust is evolving as agentic AI solutions for software develop. 

\section{Resolving Software Issues}

GitHub Co-pilot \cite{copilot} is an early approach to AI-based automation in software engineering. It attempts to provide code completions to complete code while the human programmer is typing in an Integrated Development Environment. Instead of a human programmer going to an online community such as Stackoverflow to get help in programming, it brings the knowledge from such online communities to the programmer as part of the development environment.  AI agents for coding go beyond the Co-pilot approach since the agents can exercise autonomy to make decisions and act somewhat like a (human) software engineer.

A broad structure of Large Language Model (LLM) agents for software has emerged in the past 1-2 years. An agent is autonomous and makes (informed) decisions on its own. It may be informed by invoking various external tools. In the case of LLM agents for software - the most effective solution is found if the agents work on program representations. By making the agents interpret program representations and make micro-decisions about software related tasks, we can have the agent conduct program analysis to make decisions. The program analysis can either be invoked as an external tool, or it can be embedded in the agent's actions as it interprets and navigates the program representation.

\begin{figure}
    \centering
   \includegraphics[width=0.85\linewidth]{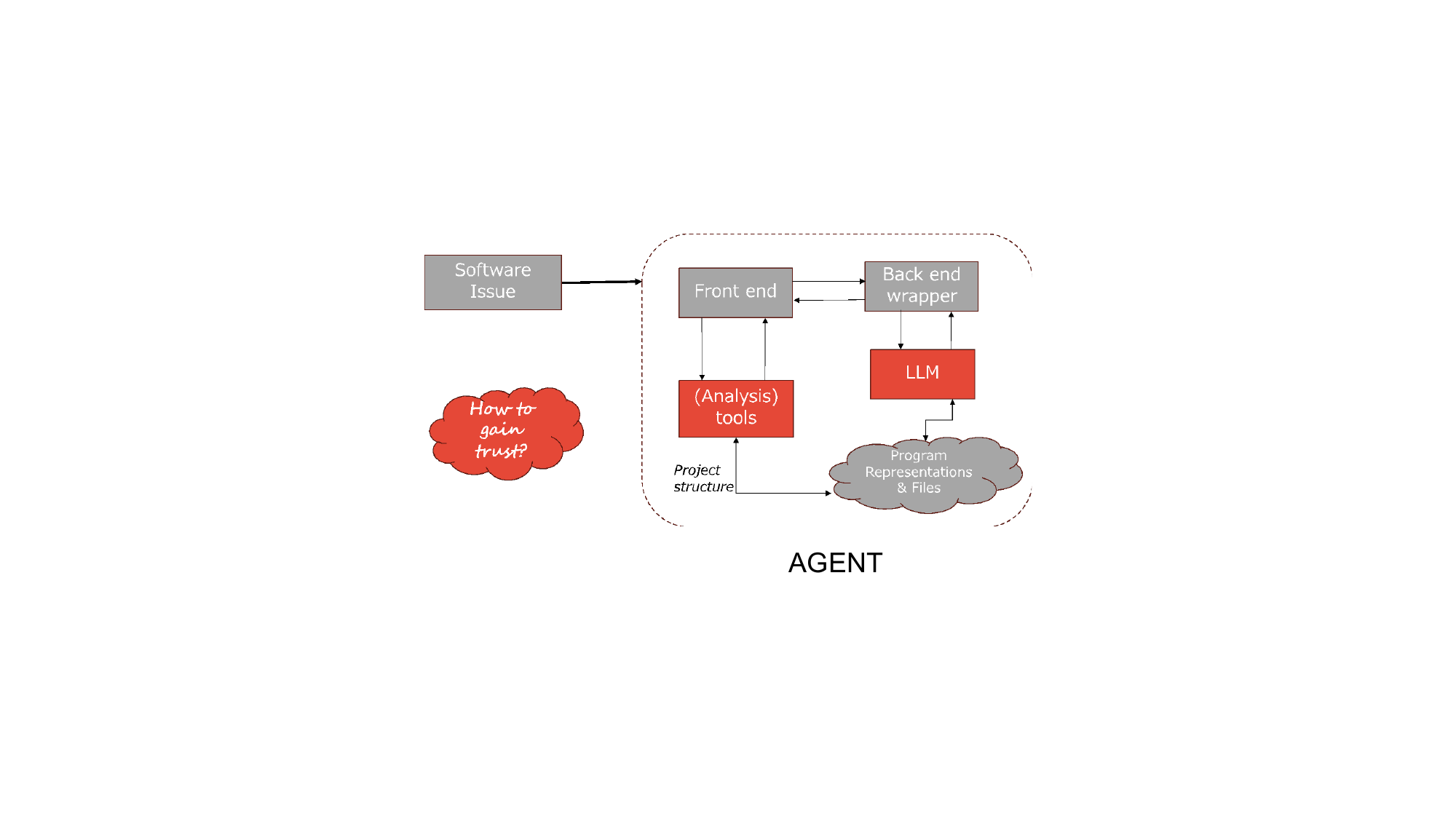}
    \caption{\em  Role of analysis in issue resolution.}
    \label{fig:issue}
\end{figure}

One of the key pivots in the growth of LLM agents for software has been the interest around automated issue resolution. A software issue is an unit of work that a (human) software engineer often needs to deal with. Such an issue is a report given to the software engineer - requesting some {\em improvement} of the software project. The improvement can focus on a bug fix, a feature addition or even a desired efficiency improvement in a core library of the software project. In recent times, open datasets of software issues such as SWE-bench \cite{swebench} have been suggested. Availability of open datasets containing open challenges have spurred the growth of agents targeting automated issue resolution in software. A schematic of an LLM agent for  conducting software issue resolution is shown in Figure \ref{fig:issue}. This schematic is given at a high level showing how an agent can invoke program analysis tools to work on program representations. At a more detailed level, an agent will use its autonomy to analyze different parts of the code-base on-the-fly in order to accomplish a software task. 

\begin{figure*}[t]
    \centering
   \includegraphics[width=0.9\linewidth]{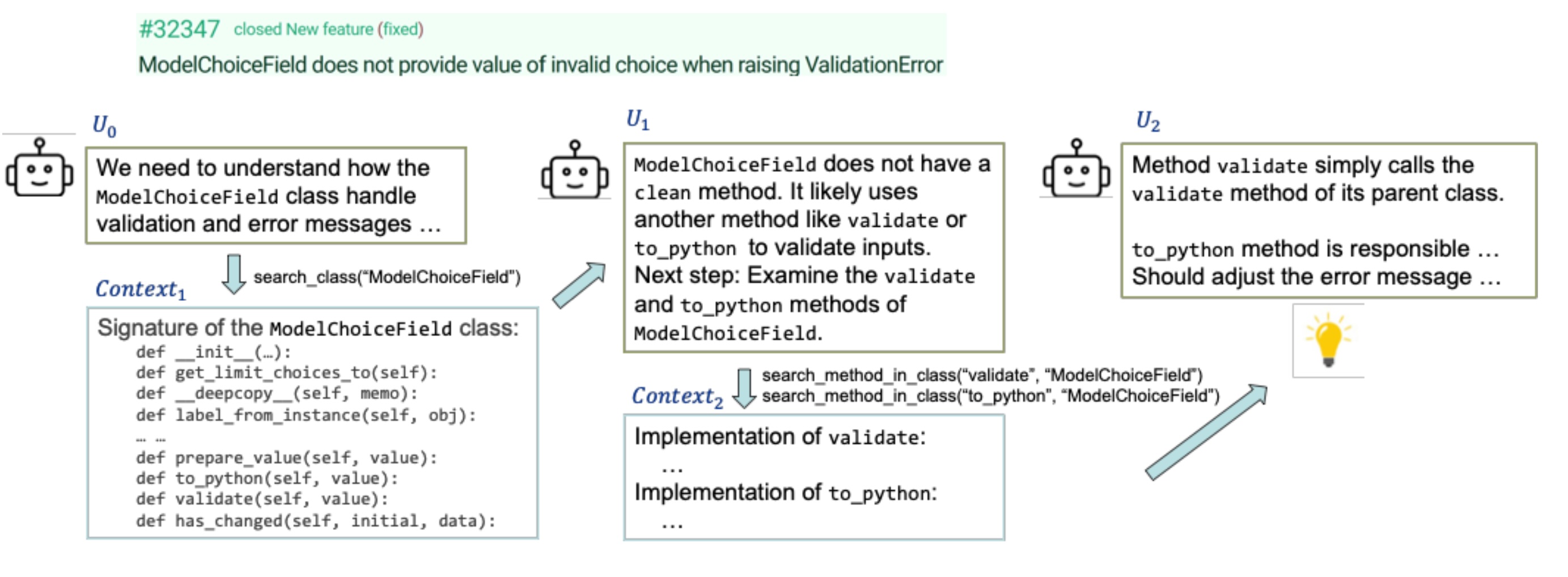}
    \caption{\em  Issue Resolution by Intent Inference in AutoCodeRover.}
    \label{fig:issue-acr}
\end{figure*}

Since the SWE-bench dataset suggested the resolution of software issues as a challenge, several AI-based agent approaches have emerged. Devin \cite{Devin24} from Cognition Labs was one of the first approaches. It suggests the use of external tools to aid the agent, thereby suggesting a first step towards AI software engineer. Devin invokes external tools such as a shell where any command can be input. Thus, the agent can autonomously work like a software engineer - inspecting files in the project and focusing on certain parts of the program to resolve software issues. The SWE agent \cite{sweagent24} provides an agentic approach that is thematically similar to Devin. It suggests the setting up of agent-computer interfaces for the agent to interact with external tools. Openhands \cite{wang2025openhands} is an open-source approach towards AI agents for code. Like a human developer it can invoke a shell, web-browser or consult Stackoverflow to modify code. 

\section{Intent Inference from Code}

All of the aforementioned approaches - Devin, SWE-agent and OpenHands - attempt to create an agentic software engineer by endowing it with the tool capabilities of the human software engineer. However, the human software engineer is also capable of understanding and reasoning about program code semantics during the software engineering process.  The process of fixing bugs or adding features in a complex code-base usually is not as simple as invoking a shell or a web-browser or looking up Stackoverflow. The human software engineer usually conducts an implicit intent inference - about what is the intended behavior of the software - and whether the current code-base is satisfying the intended behavior. This understanding of the intended behavior (and any deviation from it by the current code-base) allows the human software engineer to make code changes, thereby resolving software issues.

\paragraph{Digression: Lessons from Program Repair}
Before the arrival of AI agents, the software engineering community has studied the problem of automated program repair \cite{cacm19}. In the problem of automated program repair, the correctness criterion is given as a set of tests. The goal of automated program repair is to automatically modify a given buggy program, so that it can pass the given set of tests. However, this risks the generated patch from overfitting the given test data, it can only pass the given tests, but mostly fail tests outside the test-suite. Trying to generalize the given tests to infer intended behavior is found to be a solution to combat the overfitting problem. This lesson forms the heart of semantic program repair approaches, which use symbolic program analysis (specifically symbolic execution) for intent inference. 

\paragraph{Coming back: Learning from those lessons}
Resolving software issues includes the problem of automated program repair, but is a more general problem. Instead of given tests, the desired behavior is given as a natural language issue. Understanding the intended behavior of a software project and how it deviates from the intended behavior can help us resolve the software issue. In an agentic AI approach to building an AI software engineer - it is thus not sufficient to invoke external tools such as a shell or web-browser. Instead of the agent working on code as text, it needs to work on {\em program representations} and needs to employ program analysis for intent inference.

\paragraph{AutoCodeRover and followup}
AutoCodeRover \cite{acr24} gives a different agentic AI approach by involving program analysis. Given a software issue, its core intuition is to extract specifications (of intended behavior) from the program structure. This is achieved via code search.
Starting from the code elements (classes, methods, APIs) mentioned in the issue - it progressively searches other code elements to perform  {\em fault localization} - finding where the code changes are needed. This is then followed by a patch generation stage which considers the knowledge/specifications gained from fault localization. A follow-up work SpecRover \cite{acr25} makes the implicit intent inference in AutoCodeRover more explicit by inferring the intended behavior of each of the functions (or units off code) identified by fault localization. Figure \ref{fig:issue-acr} shows a simple example of issue resolution via intent inference. The issue being resolved is shown at the top. The code-search starts from the {\tt ModelChoiceField} class which is mentioned in the natural language issue. It then progresses by looking up the signature of this class - and then further looking into the implementation of the methdods. The AutoCodeRover agent thus iteratively refines its lower-level understanding of the intended program behavior  from the {\em program structure} such as classes / methods.

\paragraph{Use-cases and explainability} What is the benefit of the agent working on program representations? The main benefit is the inference of intended program behavior by the agent which can be used to provide explanations of the agent's actions. It enhances trust in the coding agent, and allows one to employ agents in more critical scenarios (that demand greater trust) than vibe-coding. We can also observe this from the current real-life usage of AutoCodeRover, which has now been integrated into the widely used SonarQube tool \cite{sonarqube}. SonarQube is used by enterprises for code quality and security - it provides inspection of bugs, code smells and vulnerabilities. Such bug reports generated by SonarQube can be treated as issues which are then automatically fixed by AutoCodeRover. This shows the promise of agentic AI in providing support for critical code.

\section{Verification and Validation}

Issue resolution forms the core of program improvement and maintenance of a software project.  While issue resolution and prompt based code generation have captured significant attention - the opportunities of using agentic AI for code are by no means restricted to these areas. 
With the growth of automation in software development, there will arguably be an explosion of AI-generated code. This will exacerbate the issue of verification and validation (V \& V) of the generated code. Furthermore, as the generated code gets integrated to code-bases, the interplay of legacy human written code and the generated code - will lead to subtle errors and vulnerabilities. Even if the agentic AI that produces code (fragments) is trustworthy - the integration of AI generated code can create new trust issues in software projects further down the line. There is a long history of prominent vulnerabilities caused by (human written) software. The latest prominent example is perhaps CrowdStrike outage in 2024, where the security system itself had a bug that caused worldwide outages. At the same time, AI generated code from prompts have been reported to have many vulnerabilities (see {\em e.g.}, \cite{karri}). There may be a lingering "{\em trust deficit}" in AI generated code.

Looking forward, one prominent direction thus could be to put agentic AI itself to use in combating the trust deficit from AI generated code ! Thus, given a key critical high level natural language requirement - agentic AI may help clarify the requirement by understanding its intent, as well as understanding how it is implemented at a low level in the code-base. This may lead to capabilities of security audits which have been recently studied \cite{repoaudit}. In fact, it can go beyond audits by flagging violations of the policy and fixing these violations via code changes albeit by an issue resolution agent in the back-end.

\begin{figure}
    \centering
   \includegraphics[width=\linewidth]{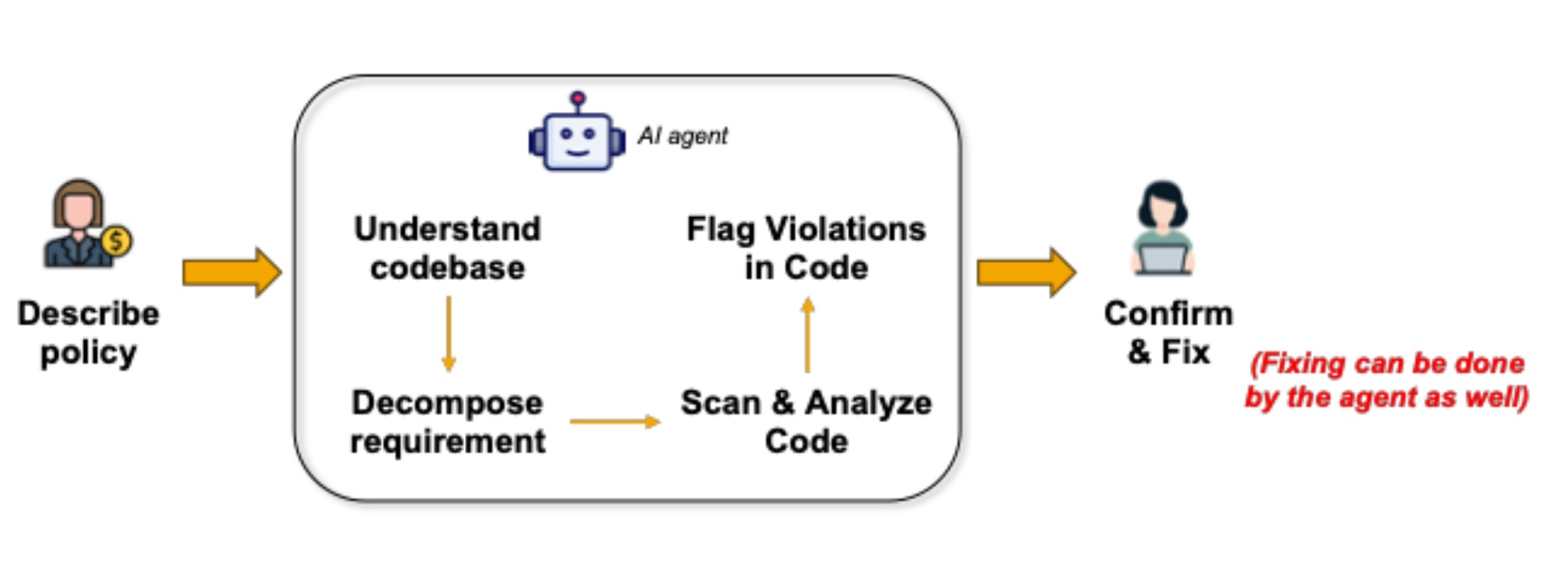}
    \caption{\em  Agentic AI for Policy Enforcement in Code.}
    \label{fig:policy}
\end{figure}

Last but not the least, agentic AI can be put to use for formal verification of code \cite{ai4x}.  Just like agents like AutoCodeRover can interpret program representations and analyze them, we can devise AI agents which can interpret formal proof representations, analyze them, and communicate with a theorem prover via an iterative refinement loop to "improve" the proof. This will also contrast with approaches like Alphaproof\footnote{https://deepmind.google/discover/blog/ai-solves-imo-problems-at-silver-medal-level/} which help prove general mathematical theorems. Instead, a program verification agent can again leverage the program structure like classes / methods to structure the proof and its intermediate lemmas. This remains an attractive future direction.

\bibliography{aaai2026}

\end{document}